\documentclass[a4paper]{article}

\usepackage{color}
\usepackage{multirow}
\usepackage{bm}
\usepackage{footnote}
\usepackage{algorithm}  
\usepackage{algorithmicx}  
\usepackage{algpseudocode}  
\usepackage{footnote}
\usepackage{hyperref}
\usepackage{INTERSPEECH2018}

\title{Improved Speaker-Dependent Separation for CHiME-5 Challenge}

\name{Jian Wu$^{1,2^*}$, Yong Xu$^3$, Shi-Xiong Zhang$^3$, Lian-Wu Chen$^2$, Meng Yu$^3$, Lei Xie$^1$, Dong Yu$^3$}

\address{
  $^1$School of Computer Science, Northwestern Polytechnical University, Xi'an, China \\
  $^2$Tencent AI Lab, Shenzhen, China \thanks{$^*$ This work was done when the first author was an intern in Tencent AI lab.} \\
  $^3$Tencent AI Lab, Bellevue, USA}
\email{\{jianwu,lxie\}@nwpu-aslp.org, \{lucayongxu,auszhang,lianwuchen,raymondmyu,dyu\}@tencent.com}

\begin{document}

\maketitle






\begin{abstract}
This paper summarizes several follow-up contributions for improving our submitted NWPU speaker-dependent system for CHiME-5 challenge, which aims to solve the problem of multi-channel, highly-overlapped conversational speech recognition in a dinner party scenario with reverberations and non-stationary noises. We adopt a speaker-aware training method by using i-vector as the target speaker information for multi-talker speech separation. With only one unified separation model for all speakers, we achieve a 10\% absolute improvement in terms of word error rate (WER) over the previous baseline of 80.28\% on the development set by leveraging our newly proposed data processing techniques and beamforming approach. With our improved back-end acoustic model, we further reduce WER to 60.15\% which surpasses the result of our submitted CHiME-5 challenge system without applying any fusion techniques.
\end{abstract}

\section{Introduction}

As the recent progress in front-end audio processing, acoustic and language modeling, automatic speech recognition (ASR) techniques are widely deployed in our daily life. However, the performance of ASR will severely degrade in challenging acoustic environments (e.g., overlapping, noisy, reverberated speech), mainly due to the unseen complicated acoustic conditions in the training. Many previous work on acoustic robustness focused on one aspect, e.g., speech separation~\cite{hershey2016deep,kolbaek2017multitalker,luo2018speaker,wang2019combining}, enhancement~\cite{xu2015regression,williamson2016complex,heymann2016neural,higuchi2017online,wang2018spatial}, dereverberation~\cite{kinoshita2016summary,wu2017reverberation,williamson2017time}, and etc. Those experiments were conducted on simulated data, which is not realistic in real applications. Recently released CHiME-5 challenge~\cite{barker2018fifth} provided a large-scale multi-speaker conversational corpus recorded via Microsoft Kinect in real home environments and targeted at the problem of distant multi-microphone conversational speech recognition. As the recordings are extremely overlapped among multiple speakers and corrupted by the reverberation and background noises, WERs reported on the dataset are fairly high. In this paper, we make several efforts based on our previously submitted speaker-dependent system~\cite{zhao2018nwpu} which ranked 3rd under unconstrained LM and 5th under constrained LM for the single device track, respectively.

The difficulties of CHiME-5 are three-fold. First, the natural conversation contains casual contents, sometimes occupied by laugh and coughing. Speaker interference is common in conversational speech as well, which causes degradation on speech recognition. Second, hardware devices, far-field wave propagation and ambient noises cause audio clipping, signal attenuation and noise corruption, respectively.
Furthermore, the lack of the clean speech for supervised training greatly limits the algorithm design and external datasets are not allowed according to the rule of CHiME-5. By considering these aspects, robust front-end processing of target speaker enhancement is critical for improving the ASR performance.

Recent studies have made great efforts in multi-channel speech enhancement~\cite{heymann2016neural,higuchi2017online,wang2018spatial,gannot2017consolidated} and most of them estimated the Time-Frequency (TF) masks that encode the speech or noise dominance in each TF unit. Deep learning based beamforming became the most popular approach since CHiME-3 and CHiME-4 challenge~\cite{barker2017third}, depending on the accurate estimation of speech covariance matrices. However, in CHiME-5 challenge, it's difficult to train the speech enhancement mask estimator and obtain accurate predictions due to the lack of the oracle clean data required by supervised training. On the other hand, there are many limitations on performing recently proposed monaural blind speech separation methods, e.g., DPCL \cite{hershey2016deep}, uPIT \cite{kolbaek2017multitalker}, because it's necessary to do speaker tracking due to the permutation issue. The number of speakers is also a prerequisite for monaural speech separation approaches, while it is infeasible in CHiME-5 challenge. However, considering that the target speaker ID is given in each utterance, we tried speaker-dependent (SD) separation in \cite{zhao2018nwpu} and Du et al. used a speaker dependent system along with a two-stage separation method in \cite{du2016ustc}. 


In this paper, we focus on single-array track and achieve significant improvement with the following contributions. First, we process data by making use of GWPE~\cite{yoshioka2012generalization}, CGMM~\cite{higuchi2017online,higuchi2016robust} and OMLSA~\cite{cohen2001speech} to further remove the interference in the non-overlapped data segments, which are used as the training target in the SD models. In \cite{zhao2018nwpu}, suffering from low-quality training targets, the system just achieved 2\% absolute reduction on WER. Second, inspired by ~\cite{zmolikova2017speaker,wang2018voicefilter,medennikov2018stc}, we incorporate i-vectors as auxiliary features, which aims to extract the target speaker. With the speaker-aware training technique, we achieve much better results using only one mask estimation model. Third, we investigate the beamforming performance, and observe that with more accurate speaker masks, generalized eigenvalue (GEV)~\cite{warsitz2007blind} beamformer performs better than minimum variance distortionless response (MVDR)~\cite{benesty2008microphone} beamformer. Finally, we report 10\% absolute WER reduction on the development set and 20\% with our improved acoustic model which is based on the factored form of time-delay neural network (TDNN-F)~\cite{povey2018semi}. Compared with the single systems submitted for CHiME-5, our proposed system outperform most of them. And compared to \cite{du2016ustc}, where a set of separation models were trained and a two-stage separation is performed, our SD method has low computational complexity apparently.

\begin{figure}[!tbp]
\label{data_flow}
\centering
\includegraphics[width=0.45 \textwidth]{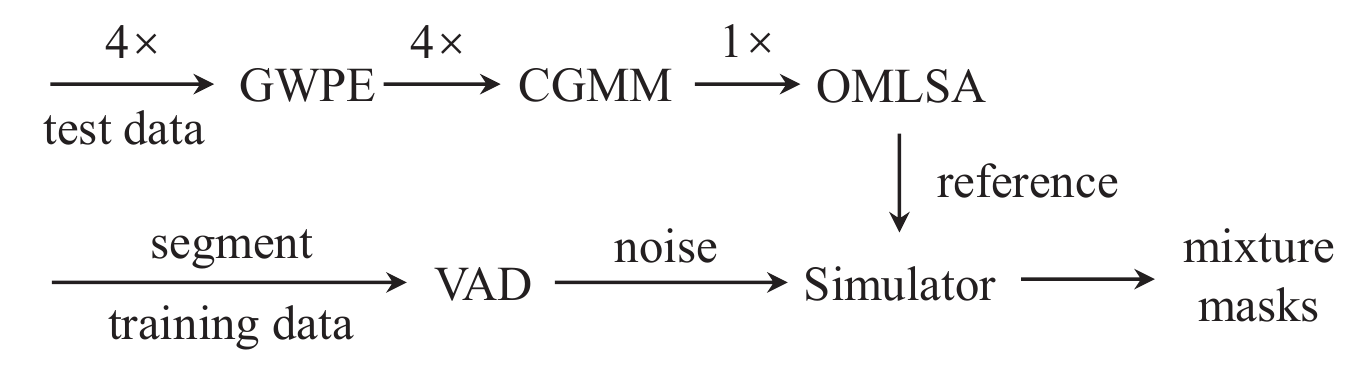}
\caption{Flow chart of data processing and simulation}
\label{block} \vspace{-0.5cm}
\end{figure}


\section{Proposed System}
In this section, we will discuss the data processing, speaker-aware training and beamforming used in our new system and indicate how we boost previously submitted speaker-dependent front-end.

\subsection{Data processing}
\label{data_proc}

In order to simulate training data for speaker-dependent models, we use non-overlapped utterances as reference, which can be segmented according to the provided annotations. However, those segments are not guaranteed to be in high signal-to-noise ratio (SNR) and may contain strong background noise, especially in the kitchen room. These issues can lead to inaccurate training targets (e.g., IRM), which may result in slow convergence and bad performance of the separation model. In order to further remove noise in those segments and improve the quality of training targets, we utilize complex Gaussian mixture model (CGMM) to estimate speech masks in a unsupervised manner and perform MVDR beamforming to suppress background noise. Following the suggestions from \cite{drude2018integrating}, GWPE is applied on multi-channel signals to reduce potential reverberations before beamforming, which are also proved to benefit ASR performance in the following experiments.

We use a two-component CGMM, i.e., speech and noise, and TF-masks are computed as the following posterior
\begin{equation}
    \lambda_{t,f}^k = \frac{p(\mathbf{y}_{t,f} | \mathbf{\Theta}_k)}{\sum_c p(\mathbf{y}_{t,f} | \mathbf{\Theta}_k)} \quad k \in \{n,s\},
\end{equation}
where $p(\mathbf{y}_{t,f} | \mathbf{\Theta}_k) = \mathcal{N}(\mathbf{y}_{t,f}| 0, \phi_{t,f}^k \mathbf{R}_{f}^k)$. Following ~\cite{heymann2016neural,higuchi2016robust}, speech and noise covariance matrices are estimated via
\begin{equation}
    \mathbf{\Phi}_f^k = \frac{1}{\sum_t \lambda_{t,f}^k} \sum_t \lambda_{t,f}^k \mathbf{y}_{t,y}\mathbf{y}_{t,y}^H \quad k \in \{n,s\},
\end{equation}
where $(\cdot)^H$ means conjugate transpose. For MVDR beamforming, steer vector $\mathbf{d}_f$ at each frequency is required, and the principal eigenvector of $\mathbf{\Phi}_f^s$ is an ideal estimation based on the fact that covariance matrices of the directional target is close to a rank-one matrix. With $\mathbf{\Phi}_f^n, \mathbf{d}_f$, weights of MVDR is computed as
\begin{equation}
\mathbf{w}_f^{\text{MVDR}} = \frac{(\mathbf{\Phi}_f^n)^{-1}\mathbf{d}_f}{\mathbf{d}_f^H(\mathbf{\Phi}_f^n)^{-1}\mathbf{d}_f}.
\end{equation}

Considering that the enhanced speech obtained by beamforming always contains residual noise, we continue to perform single-channel denoising. One typical statistical method is OMLSA \cite{cohen2001speech}, which was proposed for single-channel robust speech enhancement. Although it may introduce speech distortion, it reduces the background noise and keeps the TF regions of speech with higher energy, further improving the accuracy of target mask computation, especially in noise dominant TF bins.

As shown in Fig.1, with those processed non-overlapped segments as reference (clean) data, we perform data simulation, mask computation, network training, etc, in the following steps. 

\subsection{Speaker aware training}
\label{sa_train}
Some of recent blind speech separation methods need to know the number of speakers in the mixture and can not assign output to specific speaker properly. Here it's not suitable to use them in CHiME-5 challenge which requires to recognize the speech of target speaker in the given utterances. Under such circumstances, there are two optional methods for the front-end separation system. One is to make use of speaker information and condition the speech separation, similar to ~\cite{wang2018voicefilter}. Another one is to train a set of models for each known speaker, like the one we used in \cite{zhao2018nwpu} and also in \cite{du2016ustc}. In fact, the first one is more applicable to real scenarios because it can generalize to unseen speakers if model is well trained and it also can avoid the permutation problem at the same time.

Our motivation is to use i-vectors as speaker features to bias the prediction of the target masks. We tried two typical TF-masks, i.e, IRM and PSM, which are defined as
\begin{equation}
\begin{aligned}
    \mathbf{m}_{\text{IRM}} & = |\mathbf{s}_t|/ (|\mathbf{s}_t| + |\mathbf{n}|), \\
    \mathbf{m}_{\text{PSM}} & = |\mathbf{s}_t| \cos(\angle \mathbf{y} - \angle \mathbf{s}_t) / |\mathbf{y}|,
\end{aligned}
\end{equation}
where $\mathbf{y}, \mathbf{s}_t, \mathbf{n}$ are short-time Fourier transform (STFT) of mixture, target speaker and noise component respectively, which satisfies the equation $\mathbf{y} = \mathbf{s}_t + \mathbf{n}$. When simulating the training data, we mix target speaker with background noise as well as one or two interference speakers at various SNRs. Considering that PSM is unbounded and may be negative, we truncated its value between 0 and 1. Neural networks are trained by minimizing the mean square error
\begin{equation}
    \mathcal{L}_{\text{MSE}} = \Vert \hat{\mathbf{m}} - \mathbf{m}_t\Vert_2^2.
\end{equation}

In the training stage, for a given noisy utterance and a specific speaker, i-vectors are computed on random selected segment from target speaker's non-overlapped set, similar to \cite{wang2018voicefilter}. And during testing, we use average results instead of random one to get a robust and stable prediction. 

\subsection{Beamforming}
Beamformer is a linear spatial filter applied on microphone signals, which suppresses energy on non-target directions and produces an enhanced output. On frequency domain it could be described as
\begin{equation}
s_{t,f} = \mathbf{w}_f^H \mathbf{y}_{t,f},
\end{equation}
where $\mathbf{w}_f$ is a complex valued vector on the frequency $f$. In Section \ref{data_proc} we introduce MVDR beamforming, which is a special case of parameterized multi-channel Wiener filter (PMWF)
\begin{equation}
\mathbf{w}_f^{\text{PMWF}-\beta} = \frac{(\mathbf{\Phi}_f^n)^{-1}\mathbf{\Phi}_f^s}{\beta + \text{tr}[(\mathbf{\Phi}_f^n)^{-1}\mathbf{\Phi}_f^s]} \mathbf{u}_r
\end{equation}
with $\beta = 0$. $\mathbf{u}_r$ is a vector indicating reference microphone, which can be manually specified or chosen by the estimation of the posterior SNR~\cite{erdogan2016improved}. When $\beta = 1$, it equals to multi-channel Wiener filter (MCWF), another widely used beamforming in signal processing.

In \cite{heymann2016neural}, GEV beamformer, which is obtained by Max-SNR criterion and avoids matrix inversion in the computation, provides better results than MVDR. The beamforming filter is designed to maximize expected SNR at each frequency:
\begin{equation}
    \mathbf{w}_f^{\text{GEV}} = \arg \max_{\mathbf{w}} \frac{\mathbf{w}^H \mathbf{\Phi}_f^s\mathbf{w}}{\mathbf{w}^H \mathbf{\Phi}_f^n\mathbf{w}},
\end{equation}
which can be solved by forming a generalized eigenvalue problem with $\mathbf{\Phi}_f^s$ and $\mathbf{\Phi}_f^n$. To produce a distortionless speech signal at the beamformer output, \cite{warsitz2007blind} also provides several post-filtering algorithms to normalize GEV coefficients. In our experiments, we adopt \emph{Blind Analytical Normalization} (BAN) by default.

\section{Experiments}
\begin{table}[t]
\centering
\caption{The description of the training data}
\label{am_data}
\begin{tabular}{cccc}
\toprule
\textbf{Data ID}     & \textbf{Description}     & \textbf{Duration}           \\ \midrule
1    & worn (cleaned+sp)    & 64h$\times$3      \\ 
2    & 100k far-field (cleaned+sp)     & 39h$\times$3       \\ 
3    & reverberate on 1    & 64h$\times$3    \\ 
4    & 100k far-field (cgmm+mvdr)      & 35h      \\
5    & 100k far-field (gwpe,ch1)      & 35h      \\ 
\bottomrule
\end{tabular}
\end{table}

\begin{table}[t]
\centering
\caption{Performance of different acoustic models}
\label{am_result}
\begin{tabular}{cccc}
\toprule
\textbf{Structure}     & \textbf{Data}     & \textbf{WER\%}           \\ \midrule
baseline 9-TDNN    & 1+2    & 80.28\%     \\ 
baseline 9-TDNN    & 1+2+3  & 79.13\%     \\ 
9-TDNN+1BLSTM    & 1+2+3  & 77.15\%     \\ 
12-TDNN-F    & 1+2+3  & 70.02\%     \\ 
5CNN+9-TDNN-F    & 1+2+3  & 68.72\%     \\ 
5CNN+9-TDNN-F    & 1+2+3+4  & \textbf{68.43}\%     \\ 
Original submission \cite{zhao2018nwpu}     & - & 70.49\% \\
\bottomrule
\end{tabular}
\end{table}

\subsection{Acoustic model}
The performance of acoustic models we tuned on the development data is given in Table \ref{am_result}, with the description of the training data listed in Table \ref{am_data}. All models are trained with lattice-free maximum mutual information (LF-MMI, \cite{povey2016purely}) criterion via KALDI~\cite{povey2011kaldi} toolkit. Mel-frequency cepstral coefficients (MFCCs) and online i-vectors are adopted as input features. In addition to the training data used in official baseline ($1+2$), we include reverberated data ($3$) and enhanced data ($4+5$) processed by GWPE\footnote{\url{https://github.com/funcwj/setk/blob/master/scripts/run_gwpe.sh}} and CGMM-MVDR\footnote{\url{https://github.com/funcwj/setk/blob/master/scripts/run_cgmm.sh}}. To simulate the reverberated audio samples in $3$, we take the room impulse response (RIR) dataset released in \cite{ko2017study} but only use the portion of \emph{small room} because it has a similar room size as CHiME-5.
 
Our best configuration follows the successful practise of CNN-TDNN-F structure in our original system~\cite{zhao2018nwpu}. As can be seen in Table \ref{am_result}, our boosted version of TDNN-F acoustic model brings 12\% absolute WER reduction compared to the official TDNN, which also surpasses our previous submitted result. In the following sections, we will mainly focus on the performance of the front-end and evaluate the results with our own acoustic model (see Table \ref{tune_front_end}).

\subsection{Data processing}
\label{exp_data_processing}


\begin{figure}[t]
\centering
\includegraphics[width=0.45 \textwidth]{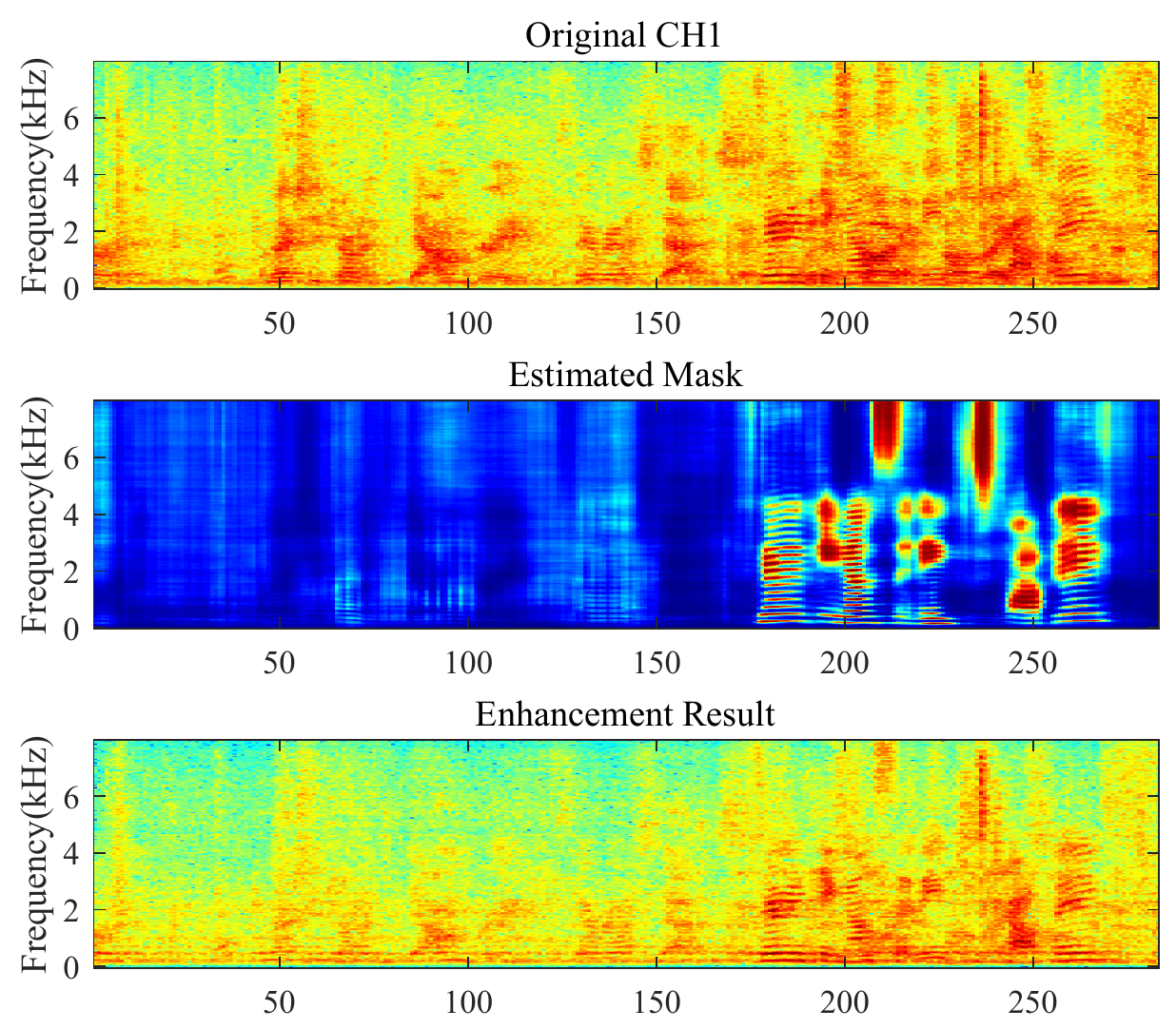}
\caption{A beamforming example and predicted target speaker masks. Although the interference speaker occupies most of the time in the utterance, the estimation of the target TF-mask is accurate. The last row plots the spectrogram of the enhancement output, where the interference speech is well suppressed.}
\label{block}
\end{figure}

\begin{table}[t]
\centering
\caption{The number of non-overlapped segments per speaker used in data simulation on the development set}
\label{nonov_segments}
\begin{tabular}{cccccccc}
\toprule
\textbf{P05} & \textbf{P06} & \textbf{P07} & \textbf{P08} & \textbf{P25} & \textbf{P26} & \textbf{P27} & \textbf{P28}          \\ \midrule
161 & 251 & 132 & 108 & 121 & 78 & 92 & 169     \\ 
\bottomrule
\end{tabular}
\end{table}

The non-overlapped segments of each speaker we used are listed in Table \ref{nonov_segments} and short segments (less than 2s) are discarded. The noise files come from non-speech intervals and a energy based VAD is used to filter out possible silence segments. Based on those processed segments and the background noise files, we simulate the data for speaker-dependent model training as depicted in Fig.1.

To demonstrate the effectiveness of the data processing discussed in Section \ref{data_proc}, we first evaluate the ASR performance of GWPE followed by CGMM-MVDR. As can be seen in Sys-5 in Table \ref{tune_front_end}, compared to the original CH1 (Sys-1), the data processing step brings 4\% absolute WER reduction. 

\begin{savenotes}
\setlength{\tabcolsep}{4pt}
\begin{table*}[t]
\centering
\caption{WER (\%) of each speaker on the development set with CNN-TDNN-F acoustic model}
\label{tune_front_end}
\begin{tabular}{cccccccccccccc}
\toprule
 \textbf{Sys} & \textbf{Input} & \textbf{Mask} & \textbf{\#Models}& \textbf{Beamformer} & \textbf{P05} & \textbf{P06} & \textbf{P07} & \textbf{P08} & \textbf{P25} & \textbf{P26} & \textbf{P27} & \textbf{P28} & \textbf{Total}          \\ \midrule
 0 & CH1-4 & \cite{zhao2018nwpu} & 8 & MVDR & 71.57 & 63.46 & 70.13 & 72.53 & 69.77 & 79.18 & 70.06 & 53.57 & 68.66 \\ \midrule
 1 & CH1 & $\times$ & - & $\times$ & 75.21 & 66.74 & 71.77 & 83.64 & 66.73 & 79.38 & 71.09 & 53.39 & 70.71 \\
 2 & GWPE-CH1 & $\times$ & - & $\times$ & 72.22 & 64.40 & 69.78 & 78.45 & 66.61 & 78.97 & 68.55 & 52.65 & 68.52 \\
 3 & CH1-4 & $\times$ & - & WDS & 72.13 & 66.09 & 69.20 & 78.10 & 67.65 & 79.93 & 68.02 & 50.78 & 68.72 \\
 4 & CH1-4 & CGMM & - & MVDR & 70.74 & 61.08 & 67.36 & 78.76 & 66.03 & 79.88 & 67.50 & 49.88 & 66.91 \\
5 & GWPE & CGMM & - & MVDR & 70.59 & 61.15 & 67.08 & 78.25 & 63.94 & 78.85 & 66.34 & 49.69 & 66.36 \\
6 & GWPE & DN & 1 &  MVDR & 70.74 & 63.15 & 68.12 & 78.63 & 63.89 & 79.34 & 64.63 & 49.06 & 66.81 \\
7 & GWPE & $\text{SD}_A$ & 8 & MVDR & 68.16 & 59.72 & 64.90 & 74.06 & 62.92 & 78.72 & 65.39 & 47.33 & 64.47 \\
8 & GWPE & $\text{SD}_B$ & 8 & MVDR & 67.57 & 60.05 & 63.92 & 72.21 & 62.13 & 76.77 & 65.08 & 46.24 & \textbf{63.75} \\
9 & GWPE & $\text{SD}_B$ & 8 & GEV & 65.95 & 59.55 & 63.57 & 65.85 & 67.34 & 76.93 & 72.62 & 50.93 & 64.43 \\
10 & GWPE & SA & 1 & MVDR & 66.72 & 59.74 & 64.53 & 71.87 & 61.36 & 76.23 & 63.95 & 46.42 & 63.37 \\
11 & GWPE & SA & 1 & GEV & 63.12 & 59.62 & 62.21 & 62.92 & 59.93 & 71.51 & 67.17 & 46.72 & \textbf{61.31} \\
12 & GWPE & SA++ & 1 & MVDR & 66.07 & 58.89 & 63.60 & 69.69 & 60.20 & 74.79 & 63.45 & 46.18 & 62.41 \\
13 & GWPE & SA++ & 1 & PMWF-1 & 65.12 & 58.03 & 63.46 & 69.66 & 60.97 & 76.12 & 64.28 & 46.09 & 62.31 \\
14 & GWPE & SA++ & 1 & GEV & 62.45 & 58.11 & 61.60 & 61.64 & 57.99 & 69.90 & 65.70 & 46.54 & \textbf{60.16} \\
\bottomrule
\end{tabular}
\end{table*}
\end{savenotes}

To evaluate the necessary of conducting single-channel denoising for each speaker, we simulate two sets of data (each with $\sim$25h) depending on whether to apply OMLSA after GWPE and CGMM-MVDR, which are denoted as $\text{SD}_A$ (without) and $\text{SD}_B$ (with), respectively. For each set, we mix target speaker with 1 or 2 interference speakers as well as background noise randomly with SDR between 0 and 10dB and SNR between -5 and 10dB. We adopt a 2$\times$TDNN-3$\times$BLSTM structure with a sigmoid output layer to estimate speaker masks and use IRM as training targets. 513-dimensional log power spectrogram features are extracted as input, with utterance level cmvn applied.

In Table \ref{tune_front_end}, we can see that with the processing step GWPE and CGMM-MVDR, $\text{SD}_A$ gives 4\% absolute WER reduction compared to official baseline (Sys-3) and including OMLSA as a further step yields better results, showing in Sys-8. Both models surpass our previous results without the data processing steps. To illustrate the necessity of speaker separation, we also train a denoising network in Sys-6 for comparison, which only predicts masks of speech instead of target speaker. Without the target information in the estimated masks, $\text{DN}$ only produces a similar result as CGMM, which inspires us to focus on separation more than enhancement or denoising in CHiME5 challenge.

\subsection{Speaker-aware training}

Our motivation is to train a speaker-independent target separation networks, which includes target speaker's embeddings as auxiliary input and outputs mask estimation of the speaker. Unfortunately, the model trained on the training set can not exceed the results mentioned above. Here we apply the idea of speaker aware training on our speaker-dependent models, as the discussed in Section \ref{sa_train}. In our experiments, we adopt the same network structure with the SD models, but concatenate i-vectors in the second layer of TDNN and the following BLSTM layer to bias the prediction of target masks. The i-vectors used here are extracted from non-overlapped segments shown in Section \ref{nonov_segments}. During the test stage, we average the i-vectors on those utterances and get one fixed embedding for each speaker. From Table \ref{tune_front_end}, $\text{SA}$ gives similar result as $\text{SD}_B$ with MVDR beamforming, but yields a significant improvement on GEV beamformer, which brings a notable WER reduction on speaker P25 $\sim$ P28. Compared to MVDR, GEV beamformer is more sensitive to TF-masks and may distort target speech and degrade ASR performance seriously if mask is estimated inaccurately.  

Based on $\text{SA}$, we utilize two strategies to further improve the performance of the speaker-aware separation and denote it as $\text{SA++}$ in the table. The first is to initialize network with a pretrained model on the training data, considering that the number of speakers and non-overlapped segments on the development is quite limited. Another one is to replace IRM with truncated PSM, which has been proved to be effective in monaural speech enhancement. We give an example in Fig.2. Row 2 shows the output masks of $\text{SA++}$ given the log power spectrogram of mixture in row 1, which masks out the interference speakers very well\footnote{More enhancement samples are available at \url{https://funcwj.github.io/online-demo/page/chime5}}. We also compare GEV with other forms of beamforming (e.g. MCWF, MVDR in \cite{erdogan2016improved}), but no better results are achieved.

Table \ref{comparation} compares the proposed system with the performance of other teams, under the circumstance that not using system combination. We get a 20\% absolute WER reduction in total compared to official result and outperform most of other teams. Although it's inferior to the USTC-iFlytek's, our system perform separation only once and has low computational complexity and model size apparently. The details on each session and location over official AM and ours are given in Table \ref{final_report}. Even based on the official backend, our SD separation front-end contributes a 10\% WER reduction, which is a significant improvement on this challenging task.

\begin{table}
\centering
\caption{Single system comparison with other teams}
\label{comparation}
\begin{tabular}{cc}
\toprule
\textbf{Team}     & \textbf{WER} (\%)    \\ \midrule
USTC-iFlytek \cite{du2016ustc} & 57.10 \\
Ours & 60.16 \\
JHU \cite{kanda2018hitachi} & 62.09 \\
Toshiba \cite{doddipatla2018toshiba} & 63.30 \\
STC \cite{medennikov2018stc} & 63.30 \\
RWTH-Paderborn \cite{kitza2018rwth} & 68.40 \\
Official \cite{barker2018fifth} & 80.28 \\
\bottomrule
\end{tabular}
\end{table}


\begin{table}
\setlength{\tabcolsep}{4pt}
\centering
\caption{WER (\%) summary on the official \& our AM}
\label{final_report}
\begin{tabular}{ccccccc}
\toprule
\textbf{AM}     & \textbf{Sess}  & \textbf{Din}  & \textbf{Kit} & \textbf{Liv}  & \textbf{Avg}  & \textbf{Total} \\ \midrule
\multirow{2}{*}{Baseline} & S02 & 70.82 & 79.79 & 62.11 & 70.26 & \multirow{2}{*}{70.46} \\
& S09 & 74.58 & 71.29 & 67.38 & 68.06 & \\
\multirow{2}{*}{Ours} & S02 & 61.58 & 70.70 & 52.58 & 60.61 & \multirow{2}{*}{\textbf{60.16}} \\
& S09 & 63.24 & 59.21 & 56.22 & 59.21 & \\
\bottomrule
\end{tabular}
\end{table}

\section{Conclusions}
In this work, we continue to optimize the performance of our speaker-dependent separation system submitted to CHiME-5 challenge. We utilize multi-channel dereverberation and enhancement algorithm, followed by single-channel denoising, to improve the quality of the training targets. To crack the data scarcity problem in CHiME-5, we apply the idea of speaker-aware training on our speaker-dependent models and reduce the number of the front-end models to one, while bringing significant ASR improvement. Experiments show that with well tuned beamforming, our system improves the ASR performance from 80.28\% official baseline to 70.46\% in terms of WER. And with our own acoustic backend, our system achieves 60.16\% WER on the development set, without using any fusion techniques.

\clearpage
\bibliographystyle{IEEEtran}

\bibliography{main}

\end{document}